\documentclass[aps,pra,twocolumn,superscriptaddress,amsmath,showpacs,nofootinbib]{revtex4-1}
\usepackage{amsmath} \usepackage{graphicx} \usepackage{amssymb}
\usepackage{epsfig} \usepackage{amsfonts}
\usepackage{longtable}

\newcommand{\etal}{{\em et al.}}
\newcommand{\tr}{\mathrm{Tr}}

\newcommand{\I}{i}

\begin{document}

\title{Measuring distances in Hilbert space by many-particle interference}

\author{Karol Bartkiewicz}\email{bark@amu.edu.pl}

\affiliation{RCPTM, Joint Laboratory of Optics of Palack\'y
University and Institute of Physics of Academy of Sciences of the
Czech Republic, 17. listopadu 12, 772 07 Olomouc, Czech Republic }
\affiliation{Faculty of Physics, Adam Mickiewicz University, PL-61-614 Pozna\'{n}, Poland}

\author{Vojt\ifmmode \check{e}\else \v{e}\fi{}ch Tr\'avn\'{\i}\ifmmode \check{c}\else \v{c}\fi{}ek }\email{vojtech.travnicek@upol.cz}
\affiliation{RCPTM, Joint Laboratory of Optics of Palack\'y
University and Institute of Physics of Academy of Sciences of the
Czech Republic, 17. listopadu 12, 772 07 Olomouc, Czech Republic }

\author{Karel Lemr}\email{k.lemr@upol.cz}
\affiliation{RCPTM, Joint Laboratory of Optics of Palack\'y
University and Institute of Physics of Academy of Sciences of the
Czech Republic, 17. listopadu 12, 772 07 Olomouc, Czech Republic }

\begin{abstract}
The measures of distances between points in a Hilbert space
are one of the basic theoretical concepts used to characterize
properties of a quantum system with respect to some etalon state.
These are not only used in studying fidelity of signal transmission 
and basic quantum phenomena but also 
applied in measuring quantum correlations, and also in quantum
machine learning. The values of quantum distance measures are very 
difficult to determine without completely reconstructing the state. 
Here we demonstrate an interferometric approach to measuring 
distances between quantum states that in some cases can outperform
quantum state tomography.  We propose
a direct experimental method to estimate such distance 
measures between two unknown two-qubit mixed states 
as Uhlmann-Jozsa fidelity (or the Bures distance),
the Hilbert-Schmidt distance, and the trace distance.
The fidelity is  estimated via the measurement of the upper 
and lower bounds of the fidelity, which
are referred to as the superfidelity and subfidelity,
respectively. Our method is based on the multiparticle 
interactions (i.e., interference) between copies of the unknown pairs of qubits. 
\end{abstract}


\maketitle

%------------------------------------------------------------------
\section{Introduction}

In classical~\cite{CoverBook} as well as in quantum~\cite{NielsenBook,BengtssonBook} communication theories the measures of distance between states quantify the accuracy of signal transmission through an imperfect communication channel. 
Here we focus on the problem of measuring or estimating three of the most popular distances by performing less measurements than required when applying full quantum state tomography. 
We discuss the problem on the example of a two-qubit states, which is of great significance to modern-day applications of quantum information processing. 
However, our method can be directly extended to be applicable to more complex systems or possibly other distance measures. 
We focus on the  analysis of nonlinear properties of two-qubit states because they play an important role in quantum protocols exploiting quantum correlations. 
Thus, establishing methods of testing  various properties of these states  is well motivated. This is especially important for photonic qubits since photons are typical carriers of quantum information used in quantum communication protocols.

The most popular signal quality quantifier is Uhlmann-Jozsa fidelity, which is also referred to as the Uhlmann transition probability.
It is commonly applied in  quantum optics, quantum information, and condensed-matter physics. 
The fidelity of two mixed quantum states represented by density matrices $\rho_1$ and $\rho_2,$ which can represent input and output states of a transmission line, was defined by Uhlmann~\cite{Uhlmann76} and Jozsa~\cite{Jozsa94} as:
\begin{equation}
F(\rho_1,\rho_2)\equiv \Big[ {\rm
Tr}\Big(\sqrt{\sqrt{\rho_1} \rho_2\sqrt{\rho_1}}\Big) \Big]^2.
\label{eq:F}
\end{equation}
This quantity is also referred to as the Uhlmann transition probability~\cite{Uhlmann76}. 
Note that the alternative definition given by Nielsen and Chuang~\cite{NielsenBook} is denoted as $\sqrt{F}$ and is sometimes also called fidelity. Some of its important properties were studied, e.g., in Refs.~\cite{Uhlmann76,Jozsa94,Mendonca08,Miszczak09}.
Fidelity can be used to construct Bures metric~\cite{Bures69}, which defines distance between density matrices of quantum states.
The Bures metric is equal to Fubini-Study metric~\cite{Fubini} when considering only pure states.
It is being used to quantify, e.g., a degree of quantum entanglement~\cite{Vedral97,Marian08} (e.g., in quantum phase transitions), a degree of polarization~\cite{polarization1,polarization2,polarization3,Gamel12}, and nonclassicality~\cite{Marian02}. 
The fidelity is related to square of Bures metric by $D_B^2(\rho_1,\rho_2) = 2[1-\sqrt{F(\rho_1,\rho_2)}]$. 

Another popular metric is the trace distance. It provides information about statistical distinguishability between two states. 
The trace distance is defined for Hermitian density matrices as
\begin{equation}
T(\rho_1,\rho_2) = \tfrac{1}{2}\tr\big(\sqrt{(\rho_1 - \rho_2)^2}\big) = \tfrac{1}{2}\sum_{i=1}|\lambda_i|,
\label{eq:TD}
\end{equation}
where $\lambda_i$ are the eigenvalues of the Hermitian matrix $(\rho_1 - \rho_2)$.  This measure due to being Euclidean  has intuitive geometric properties that can be utilized in depicting relations between quantum states (see, e.g., Ref.~\cite{tomo16})

Trace distance of two mixed states is related to the fidelity by the following inequalities
\begin{equation}
1-F(\rho_1,\rho_2) \le T(\rho_1,\rho_2) \le \sqrt{1-F(\rho_1,\rho_2)^2}.
\end{equation}
When $\rho_1$ and $\rho_2$ are pure states the upper bound on $T$ is saturated.

The last measure that we study in this paper is the Hilbert-Schmidt distance defined as
\begin{equation}
H(\rho_1,\rho_2) = \sqrt{\tr\big(\rho_1 - \rho_2\big)^2},
\label{eq:HS}
\end{equation}
which is related to trace distance via Cauchy-Schwartz inequality, i.e, $ 0 \le H(\rho_1,\rho_2) \le 2T(\rho_1,\rho_2).$

The problem that we tackle in this paper is how to efficiently measure distances in Hilbert space.
A natural solution is to perform complete quantum state tomography of $\rho_1$ and $\rho_2,$ then
to calculate a given distance measure, including the ones considered above.
As we demonstrate here, this solution cab be inefficient because in some cases it requires measuring redundant information
and the number of measurements grows exponentially with the dimension of the Hilbert space. 
Full quantum tomography in some cases can also lead to negative density matrices which require further postprocessing in order to represent physical systems. 
This problem depends on the uncertainty of the collected data and on the error-robustness of a specific tomographic protocol~\cite{tomo14,tomo16}. 
The direct calculation of fidelity via Eq.~(\ref{eq:F}) for mixed states can be a challenging task. Examples of analytic formulas
for calculating fidelity were be found, e.g., for single-qubit states~\cite{BengtssonBook} and multimode Gaussian fields~\cite{Marian12}. 
Here, we propose an interferometric approach for direct and efficient measurement of the overlaps defined as
\begin{equation}
O_n(\rho_1,\rho_2) = \tr[(\rho_1\rho_2)^n]
\label{eq:overlap}
\end{equation}
or 
$O(\rho_1,\rho_2,) = \tr(\rho_1\rho_2)$ for $n =1,$ that can be used directly to express trace distance, Hilbert-Schmidt distance, and sub- and superfidelities. 
Which are, respectively, the lower and upper bounds on the fidelity $F(\rho_1,\rho_2)$~\cite{Miszczak09,Zhou12}. 
Note that the first-order overlap $O(\rho_1,\rho_2)$ becomes purity $\chi$ for $\rho_1 = \rho_2$.  
These overlaps can be interpreted as interaction of two or four particles if the states $\rho_n$ ($n=1,2$) represent  pairs of qubits. 
Quantum circuts for measuring overlaps can be also designed using  
the method of Ekert \etal~\cite{Ekert02} based on programmable quantum networks 
with the Fredkin (controlled-SWAP) gates \cite{Fredkin82,SmolinPRA96,FiurasekPRA08,CernochPRL08,PatelSCIADV16,OnoSREP17}.
Both our and the Ekert \etal{} methods can be applied to design setups
for measuring linear and nonlinear functionals of arbitrary
states. It was shown tehoretically by Miszczak \etal~\cite{Miszczak09}
that the network method enables measuring the
first- and second-order overlaps between a pair of two-qubit
states for the estimation of their fidelity bounds. 
In this work, we apply a purely algebraic method for estimating
some second order overlaps of arbitraty two-qubit states and discuss an
experimentally-friendly linear-optical implementation.
As in Refs.~\cite{Ekert02,Miszczak09} we
assumed that we have access to copies of a given quantum
state, which can be implemented either by producing two identical
states simultaneously, or by storing the state produced earlier in
order to measure it together with the second copy of the state
available later.

We note that some experimental demonstrations of direct measurements of 
fidelity of single qubits was already reported by Du \etal~\cite{Du04}, 
Bovino \etal~\cite{Bovino05} and Adamson~\cite{Adamson08}.  
Moreover, fully-entangled fraction, which is directly related to maximum fidelity of two-qubit state with respect to a maximally entangled state, was measured in Ref.~\cite{BartkiewiczPRA17a}.
Theoretical works relevant to measuring purity and overlaps of two quantum states include,e.g., Refs.~\cite{Tanaka13,sfid}. 
These results can be applied to measuring sub- and superfidelities, which can be also measured and as described by 
\cite{Miszczak09} by following the approaches of Ekert \etal~\cite{Ekert02} and Bovino \etal~\cite{Bovino05}.
The method presened here for the experimental measurements of the first- and second-order overlaps are inspired by the method for the measurement of nonclassical correlations described in detail in Ref.~\cite{Bartkiewicz13}, which can be also used for measuring, e.g., a degree of the CHSH inequality  violation~\cite{Bartkiewicz13b,BartkiewiczPRA17a}. In contrast to previous proposals, our method is devised for easy experimental implementation on the platform of linear optics as it requires only trivial two-qubit manipulations implemented for instance by a simple beam splitter. It can also be quickly tested with hyper-entangled photons (see, e.g., Ref.~\cite{Travnicek17}) because the required operations are then a mere deterministic single-photon projections.

This article is organized as follows: In Sec.~\ref{sec:preliminaries}, we express the analyzed distance measures in terms of many-particle overlaps. In Sec.~\ref{sec:setup}, we describe efficient methods for measuring the first-and second-order overlaps. We conclude in Sec.~\ref{sec:conclusion}.

%------------------------------------------------------------------
\section{Distance measures in terms of  many-particle interference}\label{sec:preliminaries}

The density matrix of a two-qubit (quartit) system can be expressed in the Bloch representation using Einstein summation convention as

\begin{equation}\label{eq:dm}
\rho  = \tfrac{1}{4}R_{mn}\,\sigma _{m}\otimes \sigma _{n}.
\end{equation}
Here, $R_{mn}=\tr(\rho\sigma_m\otimes \sigma_n)$ are the elements of correlation matrix and $\sigma_m$, $\sigma_n$ the Pauli matrices with $m,n=0,...,3,$, and $\sigma_0 = I$ denotes the identity operator. Note that a single-qubit density matrix can be obtained after tracing out the other qubit from the two-qubit density matrix, which results in
\begin{equation}
\rho_a  = \tfrac{1}{2}R_{m0}\,\sigma _{m},\qquad \mbox{and}\qquad \rho_b  = \tfrac{1}{2}R_{0m}\,\sigma _{m}. 
\label{rho1}
\end{equation}
To make our considerations less abstract let us assume that a qubit is encoded as polarization
degree of freedom of a single, i.e.,  $\sigma_3 =|H\rangle\langle H|-|V\rangle\langle V|$, where
$H$ and $V$ correspond to horizontal and vertical polarizations, respectively.

\subsection{Fidelity, superfidelity, and subfidelity}
One can express the fidelity of single-qubit density matrices $\rho_a$ and $\rho_b$ as
\begin{equation}
F(\rho_a,\rho_b) = O(\rho_a,\rho_b) + \sqrt{S_L(\rho_a)S_L(\rho_b)},
\end{equation}
where the $S_L(\rho) = 1 - \chi(\rho)=1-O(\rho,\rho)$ is linear entropy (linear approximation to the von Neumann entropy), which can be directly measured by the method proposed in this article.

For two-qubit and higher-dimensional density matrices the situation becomes quite complicated. 
However, to estimate $F$ using a finite number of overlaps
we can use its upper and lower bounds given by Miszczak \etal in Ref.~\cite{Miszczak09}:
\begin{eqnarray}
E(\rho_1,\rho_2)\le F(\rho_1,\rho_2) \le G(\rho_1,\rho_2).
\end{eqnarray}
The lower and upper bounds are referred to as the subfidelity and superfidelity and are defined as
\begin{subequations}
\begin{align}
E(\rho_1,\rho_2)&= O(\rho_1,\rho_2) \nonumber\\&+ \sqrt{2[
O^2(\rho_1,\rho_2)-O_2(\rho_1,\rho_2)]},\label{eq:E}\\
G(\rho_1,\rho_2)&= O(\rho_1,\rho_2) + \sqrt{S_L(\rho_1)S_L(\rho_2)}.\label{eq:G}
\end{align}
\end{subequations}
To measure these bounds, the first-order overlap, the second-order overlap and the linear entropies (purities) have to be measured. 
If one or both states $\rho_1,\rho_2$ are pure, the fidelity is equal to first-order overlap $O(\rho_1,\rho_2)$.

Measuring the first-order overlaps $O(\rho_1,\rho_2)$,
$\chi(\rho_1)=O(\rho_1,\rho_1)$, and
$\chi(\rho_2)=O(\rho_2,\rho_2)$ is enough for the determination of
the superfidelity $G(\rho_1,\rho_2)$.  If it is known than one of
the states is pure, then we do not need to proceed with estimating
the subfidelity because in this case we already have all the data
needed for calculating the fidelity $F(\rho_1,\rho_2)=O(\rho_1,\rho_2)$.
In the simplest qubit case, the superfidelity and fidelity are
equivalent, i.e., $G(\rho_1,\rho_2)=F(\rho_1,\rho_2)$ and no
further work is required for estimating the fidelity. However, in
the case of quartits one also has to estimate the subfidelity
$E(\rho_1,\rho_2)$ to know in what range is the fidelity
$F(\rho_1,\rho_2)$. The only missing quantity needed for
estimating the subfidelity $E(\rho_1,\rho_2)$ is the second-order
overlap $O_2(\rho_1,\rho_2)$, which depends on the Hilbert-space
dimension of a given system (i.e., qubit or quartit) and requires
from four to eight photons.

\subsection{Trace distance}
As stated in Eq.~(\ref{eq:TD}), the trace distance equals to the half of the sum of eigenvalues for the Hermitian matrix $\Lambda=\rho_1 - \rho_2$. These eigenvalues can be calculated by using the Cayley-Hamilton theorem~\cite{C-H_theorem}, thus solving the characteristic equation $p(\Lambda)=0$, which for the $4\times4$ matrix reads
\begin{equation}
p(\Lambda) = \Lambda^4 - \tfrac{1}{2}\Pi_2 \Lambda^2 - \tfrac{1}{3}\Pi_3\Lambda  + I_4 \mathrm{det}(\Lambda) = 0,
\label{eq:char}
\end{equation}
where $\mathrm{det}(\Lambda) = \tfrac{1}{4}(\tfrac{1}{2}\Pi_2^2 - \Pi_4)$ and $\Pi_n = \tr(\Lambda)^n= \tr(\rho_1-\rho_2)^n.$
Then the roots of Eq.~(\ref{eq:char}) are the eigenvalues $\lambda$ [see Eq.~(\ref{eq:TD})] of the matrix $\Lambda$. The moments $\Pi_n$ can be decomposed into measurable overlaps simply by expanding  $\Pi_n$, i.e.,
\begin{subequations}
\begin{eqnarray}
 \Pi_1 &=& 0,\\
 \Pi_2 &=& O(\rho_1,\rho_1) + O(\rho_2,\rho_2)- 2O(\rho_1,\rho_2),\\
 \Pi_3 &=& O(\rho_1^2,\rho_1) - O(\rho_2^2,\rho_2)\nonumber\\
 &&+3[O(\rho_2^2,\rho_1)-O(\rho_1^2,\rho_2)],\\
 \Pi_4 &=& O_2(\rho_1,\rho_1)+ O_2(\rho_2,\rho_2) - 2O_2(\rho_1,\rho_2)\nonumber\\
 && +4 O(\rho_1^2,\rho_2^2) +4[O(\rho_2^3,\rho_1)-O(\rho_1^3,\rho_2)].
\end{eqnarray}
\end{subequations}
Thus, if we consider optical implementation  it is necessary to work with four to eight photons.

\subsection{Hilbert-Schmidt distance}
It is now easy to see that the Hilbert-Schmidt distance within our framework can be expressed via first-order overlaps (i.e., two-particle interference) as
\begin{equation}
H(\rho_1,\rho_2) = \sqrt{O(\rho_1,\rho_1) + O(\rho_2,\rho_2)- 2O(\rho_1,\rho_2)}.
\end{equation}
This makes it the simplest quantity to measure of the three considered metrics as in the case of two-qubit states, it requires working only with four photons.

\subsection{Two-particle overlap}
Let us first recall (see Ref.~\cite{sfid}) how the first-order overlap $O(\rho_1,\rho_2)$ [or purities $O(\rho_1,\rho_1)$ and $O(\rho_2,\rho_2)$] can be observed directly if one possesses two copies of the system. For two-qubit states the first-order overlap (purity)can be expressed as
\begin{eqnarray}
O(\rho_1,\rho_2) &=& \tfrac{1}{16}R^{(1)}_{mn}R^{(2)}_{kl}\tr
[(\sigma_m\sigma_k)\otimes(\sigma_k\sigma_l)]\nonumber\\
&=&\tfrac{1}{4} R^{(1)}_{mn}R^{(2)}_{mn}\label{eq:O12}
\end{eqnarray}
where 
\begin{subequations}
\begin{eqnarray}
R^{(1)}_{mn} &=& \tr[(\sigma_m\otimes\sigma_n)\rho_1],\\
R^{(2)}_{mn} &=& \tr[(\sigma_m\otimes\sigma_n)\rho_2].
\end{eqnarray}
\end{subequations}
To derive this relation we applied basic properties of Pauli algebra, i.e., 
\begin{equation}
\sigma_a\sigma_b = \I\varepsilon_{abc}\sigma_c+\delta_{ab}\sigma_0,\quad\mbox{and} \quad \tr(\sigma_a\sigma_b) = 2\delta_{ab},
\end{equation}
where $\I$ is imaginary unit, $a, b, c = 0,1,2,3$, $\delta_{ab}$ is the Kronecker delta. 
The Levi-Civita symbol $\varepsilon_{abc}$ is zero, if the at least two indexes are equal or $abc=0$. 
By expressing a product traces as a trace of a tensor product we obtain
\begin{eqnarray}
O(\rho_1,\rho_2)  &=& \tfrac{1}{4}\tr
[(\sigma_m\otimes\sigma_n\otimes\sigma_m\otimes\sigma_n)\nonumber
(\rho_1\otimes\rho_2)]\\
&=& \tfrac{1}{4}\tr
[(\sigma_m\otimes\sigma_m)\otimes(\sigma_n\otimes\sigma_n)\nonumber
(\rho_1\otimes\rho_2)']\\
&=& \tfrac{1}{4}\tr [(V_{a_1a_2}\otimes V_{b_1b_2})'
(\rho_1\otimes\rho_2)] ,
\end{eqnarray}
where $V=\sigma_m\sigma_m=2I -4|\Psi^-\rangle\langle\Psi^-|$,  $|\Psi^-\rangle$ is the singlet state, and $(\rho_1\otimes\rho_2)'=S_{a_2b_1}(\rho_1\otimes\rho_2)S_{a_2b_1}$, where $S_{a_2b_1}=I\otimes S\otimes I$ is unitary matrix swapping modes $b_1$ and $a_2$. 
Within this framework it is possible to introduce the Hermitian overlap operator $O$ measured on $\rho_1\otimes\rho_2,$ i.e,
\begin{equation}
O = S_{a_2b_1}V_{a_1a_2}V_{b_1b_2}S_{a_2b_1}.\label{eq:O}
\end{equation}
Measuring the purity or first-order overlp can be performed by measuring a product of two $V$ operators, which was shown in Ref.~\cite{Bartkiewicz13} can be experimentally implemented within the framework of linear optics. Alternatively, one can directly perform projections on the maximally entangled states by using wave-plates, a beam splitter and a pair of single-photon detectors (see, e.g., Refs.~\cite{tomo14,tomo16}).

\subsection{Four-particle overlap}
\begin{figure*}
\includegraphics[width=\textwidth]{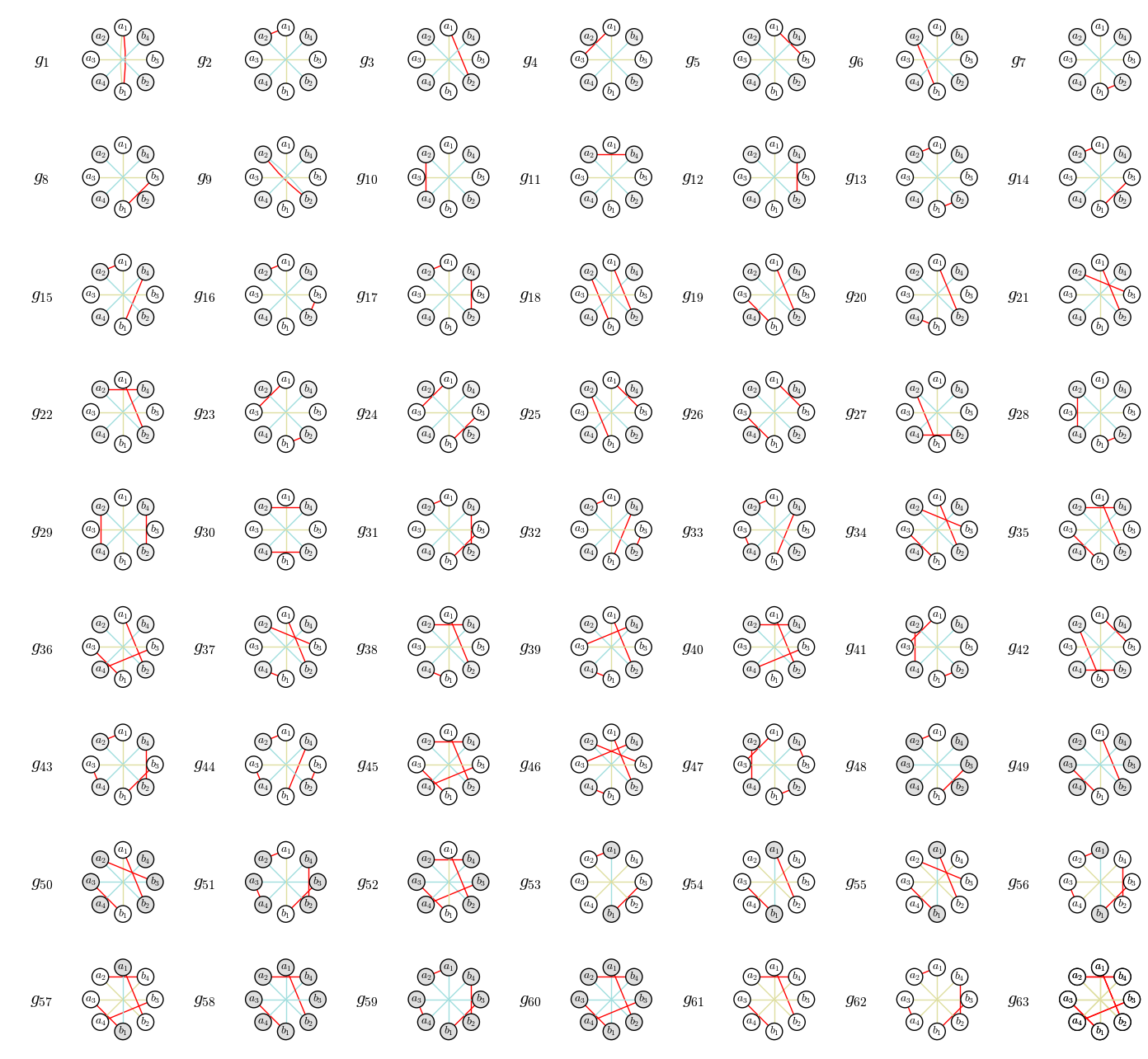}
\caption{\label{fig:all_factors}(color online) The complete set of 63 graphs representing the singlet projections needed for measuring all relevant four-particle interactions. The pairs of gray or white vertices denote the modes forming a copy of $\rho_1$ or $\rho_2$ state, respectively. The red (dark) lines mark the singlet projections (Hong-Ou-Mandel antibunching events). }
\end{figure*}

\begin{table*}[t]
\begin{ruledtabular}
\caption{\label{tab:x} Components of vector $\vec{x}$  given in terms of 63 measurement outcomes $g_n$ shown in Fig.~\ref{fig:all_factors}.}
\begin{tabular}{cc|cc|cc|cc|cc|cc|cc|cc}
$n$&$x_n$&$n$&$x_n$&$n$&$x_n$&$n$&$x_n$&$n$&$x_n$&$n$&$x_n$&$n$&$x_n$&$n$&$x_n$\\
\hline
1&$1$&23&$g_{21}$&45&$g_{1}g_{3}g_{9}$&67&$g_{38}$&89&$g_{18}^{2}$&111&$g_{1}^{3}g_{9}$&133&$g_{12}g_{13}$&155&$g_{24}^{2}$\\
2&$g_{2}$&24&$g_{22}$&46&$g_{1}g_{6}g_{9}$&68&$g_{39}$&90&$g_{9}g_{34}$&112&$g_{1}^{2}g_{18}$&134&$g_{1}g_{9}^{3}$&156&$g_{59}$\\
3&$g_{4}$&25&$g_{3}g_{6}$&47&$g_{1}g_{9}^{2}$&69&$g_{9}g_{21}$&91&$g_{45}$&113&$g_{1}g_{9}g_{26}$&135&$g_{9}^{2}g_{18}$&157&$g_{26}^{2}$\\
4&$g_{7}$&26&$g_{3}g_{9}$&48&$g_{1}g_{18}$&70&$g_{40}$&92&$g_{46}$&114&$g_{1}g_{34}$&136&$g_{1}g_{9}g_{30}$&158&$g_{60}$\\
5&$g_{8}$&27&$g_{23}$&49&$g_{1}g_{27}$&71&$g_{41}$&93&$g_{47}$&115&$g_{13}g_{24}$&137&$g_{9}g_{38}$&159&$g_{10}^{2}$\\
6&$g_{10}$&28&$g_{24}$&50&$g_{1}g_{20}$&72&$g_{10}g_{24}$&94&$g_{24}g_{29}$&116&$g_{51}$&138&$g_{13}g_{29}$&160&$g_{10}g_{12}$\\
7&$g_{12}$&29&$g_{4}g_{10}$&51&$g_{1}g_{22}$&73&$g_{12}g_{24}$&95&$g_{9}^{2}g_{26}$&117&$g_{18}g_{26}$&139&$g_{56}$&161&$g_{12}^{2}$\\
8&$g_{1}g_{9}$&30&$g_{4}g_{12}$&52&$g_{1}g_{30}$&74&$g_{4}g_{29}$&96&$g_{26}g_{30}$&118&$g_{9}g_{50}$&140&$g_{18}g_{30}$&162&$g_{9}^{2}g_{11}$\\
9&$g_{1}g_{6}$&31&$g_{25}$&53&$g_{1}^{2}g_{11}$&75&$g_{42}$&97&$g_{1}g_{5}$&119&$g_{52}$&141&$g_{1}g_{55}$&163&$g_{9}g_{54}$\\
10&$g_{1}g_{11}$&32&$g_{26}$&54&$g_{2}g_{13}$&76&$g_{9}g_{25}$&98&$g_{48}$&120&$g_{9}g_{11}$&142&$g_{57}$&164&$g_{10}g_{29}$\\
11&$g_{1}g_{3}$&33&$g_{5}g_{9}$&55&$g_{7}g_{13}$&77&$g_{9}g_{26}$&99&$g_{2}g_{4}$&121&$g_{53}$&143&$g_{4}^{2}$&165&$g_{12}g_{29}$\\
12&$g_{1}^{2}$&34&$g_{5}g_{11}$&56&$g_{31}$&78&$g_{11}g_{26}$&100&$g_{4}g_{7}$&122&$g_{2}g_{10}$&144&$g_{4}g_{8}$&166&$g_{11}g_{30}$\\
13&$g_{13}$&35&$g_{27}$&57&$g_{32}$&79&$g_{5}g_{9}^{2}$&101&$g_{49}$&123&$g_{7}g_{10}$&145&$g_{8}^{2}$&167&$g_{61}$\\
14&$g_{14}$&36&$g_{6}g_{9}$&58&$g_{33}$&80&$g_{5}g_{30}$&102&$g_{2}g_{8}$&124&$g_{54}$&146&$g_{1}^{2}g_{5}$&168&$g_{9}^{4}$\\
15&$g_{15}$&37&$g_{28}$&59&$g_{3}g_{18}$&81&$g_{8}g_{29}$&103&$g_{7}g_{8}$&125&$g_{2}g_{12}$&147&$g_{1}g_{49}$&169&$g_{9}^{2}g_{30}$\\
16&$g_{16}$&38&$g_{7}^{2}$&60&$g_{6}g_{18}$&82&$g_{1}^{2}g_{9}^{2}$&104&$g_{1}^{3}$&126&$g_{7}g_{12}$&148&$g_{4}g_{24}$&170&$g_{9}g_{55}$\\
17&$g_{17}$&39&$g_{8}g_{10}$&61&$g_{9}g_{18}$&83&$g_{1}g_{9}g_{18}$&105&$g_{1}g_{26}$&127&$g_{9}^{3}$&149&$g_{8}g_{24}$&171&$g_{29}^{2}$\\
18&$g_{2}^{2}$&40&$g_{8}g_{12}$&62&$g_{34}$&84&$g_{1}g_{38}$&106&$g_{2}g_{24}$&128&$g_{9}g_{30}$&150&$g_{5}g_{26}$&172&$g_{62}$\\
19&$g_{2}g_{7}$&41&$g_{9}^{2}$&63&$g_{35}$&85&$g_{1}^{2}g_{30}$&107&$g_{7}g_{24}$&129&$g_{2}g_{29}$&151&$g_{58}$&173&$g_{30}^{2}$\\
20&$g_{18}$&42&$g_{29}$&64&$g_{36}$&86&$g_{13}^{2}$&108&$g_{4}g_{13}$&130&$g_{7}g_{29}$&152&$g_{1}^{4}$&174&$g_{63}$\\
21&$g_{19}$&43&$g_{30}$&65&$g_{9}g_{19}$&87&$g_{43}$&109&$g_{50}$&131&$g_{10}g_{13}$&153&$g_{1}^{2}g_{26}$\\
22&$g_{20}$&44&$g_{1}^{2}g_{9}$&66&$g_{37}$&88&$g_{44}$&110&$g_{8}g_{13}$&132&$g_{55}$&154&$g_{1}g_{50}$
\end{tabular}
\end{ruledtabular}
\end{table*}
Here we describe the main result of our paper, i.e., the second order overlap for four-particle interference that are necessary to measure subfidelity $E$ given in (\ref{eq:E}) and trace distance $T$
defined in (\ref{eq:TD}). 
Note that, we can use the same reasoning as in this case to discuss lower-order interactions by preparing one or more of the particles in a completely mixed state.
To develop to only multi-particle Hong-Ou-Mandel interference based method of measuring the the second order overlap we  have applied the procedure described below. 
The calculations are nontrivial as they require utilizing a number of algebraic properties and because of the complexity of the problem we also used a computer algebra system~\cite{sage}. 
Let us start with expressing a product of two density matrices as
\begin{equation}
\rho_1\rho_2 = \tfrac{1}{16} R^{(1)}_{mn} R^{(2)}_{kl} (\sigma_m\sigma_k)^{(1)} \otimes (\sigma_n\sigma_l)^{(2)}.
\end{equation}
Now, the second order overlap can be expressed as
\begin{equation}
\tr{}(\rho_1\rho_2)^2 = \tr{}[S (\rho_1\rho_2)^{(12)}\otimes (\rho_1 \rho_2)^{(34)}],
\end{equation}
where the shift operator reads
\begin{equation}
S = S_{23}S_{34}S_{12}S_{23}.
\end{equation}
Thus, using the cyclic property of trace we have
\begin{eqnarray}\nonumber
\tr{}(\rho_1\rho_2)^2 &=& 2^{-8}R_{mn}^{(1)}R_{kl}^{(2)} R_{xy}^{(1)}R_{rs}^{(2)} \tr{}[S'   (\sigma_m\sigma_k)^{(1)} \\&&\otimes(\sigma_x\sigma_r)^{(2)}\otimes(\sigma_n\sigma_l)^{(3)}(\sigma_r\sigma_s)^{(4)}],
\end{eqnarray}
where $S' = S_{34}S_{12},$ $S_{12}=1-2P^-_{12},$  $S_{23}=1-2P^-_{23},$ and $P^- = \tfrac{1}{4}(1-\sigma_i\otimes\sigma_i).$ Further transforming the shift operator $S'$ results in
\begin{equation}
S' = \tfrac{1}{4}(1-\sigma_i^{(1)}\otimes \sigma_i^{(2)} - \sigma_j^{(3)}\otimes \sigma_j^{(4)} + \sigma_i^{(1)}\otimes \sigma_i^{(2)}\otimes \sigma_j^{(3)}\otimes \sigma_j^{(4)}).
\end{equation}
Hence, we can express the overlap as
\begin{eqnarray}\nonumber
\tr{}(\rho_1\rho_2)^2 &=&2^{-10}R_{mn}^{(1)}R_{kl}^{(2)} R_{xy}^{(1)}R_{rs}^{(2)} \tr{} [(1-\sigma_i^{(1)}\otimes \sigma_i^{(2)} \\ \nonumber
&&- \sigma_j^{(3)}\otimes \sigma_j^{(4)}+ \sigma_i^{(1)}\otimes \sigma_i^{(2)}\otimes \sigma_j^{(3)}\otimes \sigma_j^{(4)})\\&&\times
 (\sigma_m\sigma_k)^{(1)} \otimes(\sigma_x\sigma_r)^{(2)}\otimes(\sigma_n\sigma_l)^{(3)}(\sigma_r\sigma_s)^{(4)}]\nonumber\\
\end{eqnarray}
or equivalently as
\begin{equation}
\tr{}(\rho_1\rho_2)^2 = R_{mn}^{(1)}R_{kl}^{(2)} R_{xy}^{(1)}R_{rs}^{(2)}   [A_1-A_2-A_3+A_4]
\label{eq:general}
\end{equation}
where the sate-independent tensors read as
\begin{subequations}
\begin{eqnarray}
A_1&=& 2^{-10}\delta_{mk} \delta_{xr} \delta_{nl} \delta_{ys},\\
A_2&=&  2^{-6}\left( \delta^{(3)}_{rx} \delta^{(3)}_{mk} + \delta^{(3)}_{mx} \delta^{(3)}_{rk} -\delta^{(3)}_{mr} \delta^{(3)}_{kx}\right.\nonumber\\
             & & + \delta^{(3)}_{mr} \delta_{k0}\delta_{x0} + \delta^{(3)}_{kx} \delta_{r0}\delta_{m0}  + \delta^{(3)}_{kr} \delta_{m0}\delta_{x0}\nonumber\\
             & &\left. + \delta^{(3)}_{mk} \delta_{r0}\delta_{x0} + \delta^{(3)}_{rx} \delta_{k0}\delta_{m0}  + \delta_{k0} \delta_{m0}\delta_{x0}\delta_{r0}\right),\\
A_3&=&  2^{-6}\left( \delta^{(3)}_{sy} \delta^{(3)}_{nl} + \delta^{(3)}_{ny} \delta^{(3)}_{sl} -\delta^{(3)}_{ns} \delta^{(3)}_{ly}\right.\nonumber\\
             & & + \delta^{(3)}_{ns} \delta_{l0}\delta_{y0} + \delta^{(3)}_{ly} \delta_{s0}\delta_{n0}  + \delta^{(3)}_{ls} \delta_{n0}\delta_{y0}\nonumber\\
             & &\left. + \delta^{(3)}_{nl} \delta_{s0}\delta_{y0} + \delta^{(3)}_{sy} \delta_{l0}\delta_{n0}  + \delta_{l0} \delta_{n0}\delta_{y0}\delta_{s0}\right),\\
A_4 &=& 2^{-10}(\delta^{(3)}_{im}\delta_{k0} + \delta^{(3)}_{ik}\delta_{m0} + \delta^{(3)}_{mk}\delta_{i0} + \delta_{i0}\delta_{m0}\delta_{k0}\nonumber\\
    &&+ i\epsilon_{mki}) \times (\delta^{(3)}_{ix}\delta_{r0} + \delta^{(3)}_{ir}\delta_{x0} + \delta^{(3)}_{xr}\delta_{i0} \nonumber\\
    &&+ \delta_{i0}\delta_{x0}\delta_{r0}+ i\epsilon_{xri})\times (\delta^{(3)}_{jn}\delta_{l0} + \delta^{(3)}_{jl}\delta_{n0} + \delta^{(3)}_{nl}\delta_{j0}  \nonumber\\
    &&+ \delta_{j0}\delta_{n0}\delta_{l0}+ i\epsilon_{nlj})\times (\delta^{(3)}_{jy}\delta_{s0} + \delta^{(3)}_{js}\delta_{y0} + \delta^{(3)}_{sy}\delta_{j0} \nonumber\\ 
    &&+ \delta_{j0}\delta_{s0}\delta_{y0}+ i\epsilon_{ysj}).
\end{eqnarray}
\end{subequations}
The resulting expressions are rather complex as they describe 2-, 3-, and 4-particle interactions. However, due to the properties of Pauli algebra, we can express the final outcome as a polynomial of 2-particle interactions that can be implemented in an optical setup by Hong-Ou-Mandel antibunching events.
Remarkably, we can simplify the resulting expression for the second order overlap so that it is  given by a small number of terms by observing that matrix multiplication of the form $R^{(1)}_{mn}R^{(2)}_{nk}$ corresponds to measuring singlet projection $R^{(1)}_{mn}R^{(2)}_{nk}=\tr{}[(\rho_1 \otimes \rho_2) \,\sigma^{(1a)}_m\otimes \sigma^{(2b)}_k \otimes (1-4P^{-}_{1b2a})]=\tr{}[(\rho_1 \otimes \rho_2)\,\sigma^{(1a)}_m\otimes \sigma^{(2b)}_k ]-4\tr{}[(\rho_1 \otimes \rho_2) \,\sigma^{(1a)}_m\otimes \sigma^{(2b)}_k \otimes P^{-}_{1b2a}].$ This expression for  $R^{(1)}_{mn}R^{(2)}_{nk}$
can be represented graphically as presented in Fig.~\ref{fig:all_factors}. In similar manner (with help of computer algebra program~\cite{sage}), by using Eq.~(\ref{eq:general}) the expression for the four-particle overlaps can be written as 
\begin{subequations}
\begin{eqnarray}
O(\rho_1^3,\rho_1)&=&\tr(\rho_1\rho_1\rho_1\rho_1)=\vec\theta^{(1111)}\cdot\vec x,\\
O(\rho_1^3,\rho_2)&=&\tr(\rho_1\rho_1\rho_1\rho_2)=\vec\theta^{(1112)}\cdot\vec x,\\
O(\rho_1^2,\rho_2^2)&=&\tr(\rho_1\rho_1\rho_2\rho_2)=\vec\theta^{(1122)}\cdot\vec x,\\
O_2(\rho_1,\rho_2)&=&\tr(\rho_1\rho_2\rho_1\rho_2)=\vec\theta^{(1212)}\cdot\vec x,\\
O(\rho_1,\rho_2^3)&=&\tr(\rho_1\rho_2\rho_2\rho_2)=\vec\theta^{(1222)}\cdot\vec x,\\
O(\rho_2,\rho_2^3)&=&\tr(\rho_2\rho_2\rho_2\rho_2)=\vec\theta^{(2222)}\cdot\vec x,
\end{eqnarray}
\end{subequations}
where $\vec\theta^{(n)}$ for $n=1111,1112,1122,1212,1222,2222$ are vectors of state-independent coefficients (for their explicit form Tab.~\ref{tab:theta}) and components of $\vec{x}$ are given in Tab.~\ref{tab:x}. Note that in the second overlap used to 
estimate subfidelity is given by $\vec\theta^{(1212)}$.
Remarkably, the moments $\Pi_n$ can be also expressed as similar dot-products, i.e.,
\begin{subequations}
\begin{eqnarray}
\Pi_2&=&\tr(\rho_1-\rho_2)^2=\vec\beta^{(2)}\cdot\vec x,\label{eq:pi2}\\
\Pi_3&=&\tr(\rho_1-\rho_2)^3=\vec\beta^{(3)}\cdot\vec x,\\
\Pi_4&=&\tr(\rho_1-\rho_2)^4=\vec\beta^{(4)}\cdot\vec x.
\end{eqnarray}
\end{subequations}
All three $\vec\beta^{(n)}$ vectors can be found in Tab.~\ref{tab:beta}. 
To obtain these final expressions we have summed the equivalent graphs and factorized the remaining unique graphs (see elements of $\vec{x}$ in Tab.~\ref{tab:x} and $g_n$ shown in Fig.~\ref{fig:all_factors}). Note that the number of measurements $g_n$ needed to determine $\vec{x}$ is $63$. The number of measurements (projections) required to perform full two-qubit tomography for two arbitrary two-qubit states (30 projections per state).

%==================================================================
\section{Designing a multiparticle interferometer} \label{sec:setup}

\subsection{Hilbert-Schmidt distance}
Our results can be used in practice. In an experiment designed to measure Hilbert-Schmidt metrics we would use two copies of both states. The original state is encoded as two-photon polarization, while the copy is encoded into spatial degree of freedom. 
By using this approach we are able to implement all the measurements required for determining $\Pi_2,$ which correspond to 
$g_n$ for $n=2,4,7,8,10,12,13,24,29.$ This is because $\vec{x}$ measurements corresponding to nonzero  elements of $\vec\beta^{(2)}$ can be expressed as products of other measurements making the number of the prime measurements as low as $9$. 
For $\vec\beta^{(2)}$ the nonzero elements are
$\beta^{(2)}_{2}=4,$
$\beta^{(2)}_{3}=-2,$
$\beta^{(2)}_{4}=4,$
$\beta^{(2)}_{5}=-2,$
$\beta^{(2)}_{6}=-2,$
$\beta^{(2)}_{7}=-2,$
$\beta^{(2)}_{13}=-8,$
$\beta^{(2)}_{28}=4,$
$\beta^{(2)}_{42}=4.$
Note that it is sufficient to design an interferometer corresponding to the most complex graphs $g_{13},$ $g_{24},$ and $g_{29}$ as the less complex graphs are measured if the singlet projection is replaced by identity operation (i.e., intensity measurement). Hence, $\Pi_2$ can be measured in a 6-photon-pair interferometric configuration $i_n$ for $n=1$ associated with the projective measurements $g_{13},g_{24},g_{29},$ where 6 photon pairs are used and $V$ operators are measured instead of singlet projections. At the same time, while using only 6 photon pairs it is possible to access only 6 out of 30 parameters of density matrices. This demonstrates the superior performance (in terms of the number of measured  parameters) of the interferometric method with respect to the tomographic approach.

\subsection{Fidelity, subfidelity, and superfidelity}
We can apply our results to design an experiment aimed at measuring subfidelity $G$. As it follows from Eq.~(\ref{eq:G}) in addition to measuring first-order overlap, the second order overlap $O_2(\rho_1,\rho_2)$ needs to be measured. Within our framework of singlet projections, superfidelity can be measured as described in Ref.~\cite{sfid}. However, the method for measuring subfidelity  presented in 
in Ref.~\cite{sfid}, does not utilize simple antibunching events and may be problematic in implementation. Here, present an alternative solution which is free of this shortcoming. The set of measurements needed to estimate $O_2$ 
corresponds to  antibunching events shown in Fig.~\ref{fig:all_factors} , where $g_n$ for $n=1,2,3,4,5,6,7,8,9,10,11,12,13,15,16,18,19,20,21,22,$ $24,25,26,27,29,30,31,32,33,34,35,37,38,39,41,42,43,$ $44,45,46,47,$ This makes the total number of required projections equal to 41.  However, after closer examination, we see that all the required quantities can be measured in in 4-photon-pair interferometric configurations $i_n$ for $n=1,2,3,4,5,6,7,8,9$ associated with the projective measurements in the following way:
$i_1 = g_{10}g_{34},$
$i_2 = g_{1}g_{38},$
$i_3 = g_{13}g_{18},$
$i_4 = g_{24}g_{29},$
$i_5 = g_{26}g_{30},$
$i_6 = g_{43},$
$i_7 = g_{44},$
$i_8 = g_{45},$
$i_9 = g_{46},$
$i_{10} = g_{47},$ where 20 photon pairs are used and $V$ operators are measured instead of singlet projections. In case of full quantum state tomography, we would have measured 20 of 60 required projective measurements (i.e., 20 of 30 required tensor products of Pauli operators) while using the same number of photon pairs. Thus, measuring subfidelity can be performed more efficiently with interferometric method, without recourse to full quantum-state tomography.

\subsection{Trace distance}
For measuring trace distance $T(\rho_1,\rho_2)$ with our method we would measure $\Pi_n$ for $n=2,3,4,$  then calculate the eigenvalues of $\Lambda$ as given by Eq.~(\ref{eq:char}) by replacing $\Lambda$ with 
with a variable for which we solve the characteristic equation. Finally, we would use definition (\ref{eq:TD}). 
The projections required for measuring $\Pi_3,$ 
are $g_n$ for $n=1,3,5,6,9,11,14,15,16,17,18,19,20,21,22,23,25,26,27,$ $28,30,34,38,48,49,50,53,54,55$ 
and additionally 
$n=2,4,7,8,10,12,13,24,29,43,44,45,46,47,51,52,56,57,$ $59,60,62,63$ for $\Pi_4$. The total amount of projections required to determine $T$ via moments is 51 (using 104 photon pairs). This number is larger than 60 protective measurements required for quantum state tomography. 
Thus, measuring $T$ with interferometric method is more challenging than applying two-qubit tomography.

\begin{table}[t]
\begin{ruledtabular}
\caption{\label{tab:beta} The nonzero components of vectors $\beta^{(n)}$ for $n=2,3,4$ (used for calculating moments $\Pi_n$) corresponding to components of vector $\vec{x}$ described in Tab.~\ref{tab:x}.}
\begin{tabular}{cc|cc|cc|cc|cc|cc}
$n$&$\beta_n^{(2)}$&$n$&$\beta_n^{(3)}$&$n$&$\beta_n^{(3)}$&$n$&$\beta_n^{(4)}$&$n$&$\beta_n^{(4)}$&$n$&$\beta_n^{(4)}$\\
\hline
$2$&$4$&$9$&$-6$&$61$&$-24$&$18$&$8$&$91$&$32$&$119$&$-32$\\
$3$&$-2$&$10$&$6$&$62$&$-24$&$19$&$16$&$92$&$16$&$122$&$-8$\\
$4$&$4$&$11$&$-6$&$67$&$24$&$29$&$4$&$93$&$-16$&$123$&$-8$\\
$5$&$-2$&$14$&$-6$&$77$&$12$&$30$&$4$&$94$&$24$&$125$&$-8$\\
$6$&$-2$&$15$&$6$&$97$&$6$&$38$&$8$&$95$&$8$&$126$&$-8$\\
$7$&$-2$&$16$&$-6$&$98$&$6$&$39$&$4$&$96$&$-8$&$129$&$16$\\
$13$&$-8$&$17$&$6$&$101$&$-6$&$40$&$4$&$99$&$-8$&$130$&$16$\\
$28$&$4$&$21$&$6$&$104$&$4$&$54$&$-32$&$100$&$-8$&$131$&$16$\\
$42$&$4$&$22$&$-6$&$105$&$-12$&$55$&$-32$&$102$&$-8$&$133$&$16$\\
 & &$23$&$6$&$109$&$8$&$72$&$-8$&$103$&$-8$&$134$&$5\tfrac{1}{3}$\\
 & &$24$&$-6$&$120$&$-6$&$73$&$-8$&$106$&$16$&$135$&$-16$\\
 & &$26$&$6$&$121$&$-6$&$74$&$-8$&$107$&$16$&$136$&$-16$\\
 & &$27$&$-6$&$124$&$6$&$81$&$-8$&$108$&$16$&$137$&$32$\\
 & &$31$&$6$&$127$&$-4$&$82$&$-8$&$110$&$16$&$138$&$-48$\\
 & &$33$&$-6$&$128$&$12$&$83$&$32$&$111$&$5\tfrac{1}{3}$&$139$&$32$\\
 & &$35$&$-6$&$132$&$-8$&$84$&$-32$&$112$&$-16$&$140$&$16$\\
 & &$36$&$6$& & &$85$&$8$&$113$&$-16$&$141$&$10\tfrac{2}{3}$\\
 & &$37$&$6$& & &$86$&$48$&$114$&$32$&$142$&$-32$\\
 & &$44$&$-12$& & &$87$&$-16$&$115$&$-48$&$143$&$2$\\
 & &$47$&$12$& & &$88$&$-16$&$116$&$32$& & \\
 & &$48$&$24$& & &$89$&$-16$&$117$&$16$& & 
 \end{tabular}
\end{ruledtabular}
\end{table}

%------------------------------------------------------------------
\section{Conclusions \label{sec:conclusion}}

In this paper we put forward a direct method for measuring the lower bound (i.e., subfidelity~\cite{Miszczak09}) of the Uhlmann-Jozsa fidelity (equivalent to the Bures distance), Hilbert-Schmidt distance and trace distance for arbitrary unknown mixed two-qubit states. 
Our proposal of a experimentally-friendly method  for direct measuring the second-order overlaps of two arbitrary two-qubit states enables the qualitative determination of their similarity. In particular, we demonstrated that while having access simultaneously to 12 photons, we could measure Hilbert-Schmidt distance directly. To this date, experiments utilizing interference of up to 16 photons in linear optical circuts have been reported, see, e.g., \cite{Wang18}. 
This lets us believe, that our method can be implemented experimentally. Distances between states can be utilized as cost functions for optimization problems solved in quantum machine learning \cite{Cai15,Biamonte17}. Thus, we believe that our results can be useful in this context.

For Hilbert-Schmidt distance the number of required projective measurements is 10, which is smaller than 60 required for full quantum state tomography of both two-qubit states. In case of trace distance the number of the required projective measurements is much larger. As discussed in Sec.~\ref{sec:setup}, while analyzing both subfidelity and Hilbert-Schmidt distance, many of the measurements can be performed simultaneously. However, the example of trace distance shows that many-particle interference based method does not always outperform quantum state tomography.

\begin{table*}[t]
\begin{ruledtabular}
\caption{\label{tab:theta} The nonzero components of vectors $\theta^{(n)}$ for $n=1111,2222,1112,1222,1212,1122$ (used to calculate four-particle overlaps)
corresponding to components of vector $\vec{x}$ described in Tab.~\ref{tab:x}.}
\begin{tabular}{cc|cc|cc|cc|cc|cc|cc|cc|cc|cc|cc}
$n$&$\theta_n^{(1111)}$&$n$&$\theta_n^{(2222)}$&$n$&$\theta_n^{(1112)}$&$n$&$\theta_n^{(1112)}$&$n$&$\theta_n^{(1222)}$&$n$&$\theta_n^{(1222)}$&$n$&$\theta_n^{(1212)}$&$n$&$\theta_n^{(1212)}$&$n$&$\theta_n^{(1122)}$&$n$&$\theta_n^{(1122)}$&$n$&$\theta_n^{(1122)}$\\
\hline
$1$&$1$&$1$&$1$&$1$&$1$&$99$&$2$&$1$&$1$&$122$&$2$&$1$&$1$&$57$&$-8$&$1$&$1$&$34$&$2$&$71$&$-4$\\
$3$&$-4$&$6$&$-4$&$2$&$-2$&$100$&$2$&$2$&$-2$&$123$&$2$&$2$&$-4$&$58$&$-8$&$2$&$-2$&$35$&$-2$&$72$&$-2$\\
$5$&$-4$&$7$&$-4$&$3$&$-2$&$101$&$-2$&$4$&$-2$&$124$&$-2$&$4$&$-4$&$59$&$-8$&$3$&$-1$&$36$&$2$&$73$&$-2$\\
$12$&$-4$&$41$&$-4$&$4$&$-2$&$102$&$2$&$6$&$-2$&$125$&$2$&$8$&$-4$&$60$&$-8$&$4$&$-2$&$37$&$2$&$74$&$-2$\\
$28$&$4$&$42$&$4$&$5$&$-2$&$103$&$2$&$7$&$-2$&$126$&$2$&$10$&$4$&$63$&$8$&$5$&$-1$&$39$&$1$&$77$&$-2$\\
$32$&$4$&$43$&$4$&$8$&$-2$&$104$&$1\tfrac{1}{3}$&$8$&$-2$&$127$&$1\tfrac{1}{3}$&$13$&$4$&$65$&$-8$&$6$&$-1$&$40$&$2$&$78$&$-2$\\
$97$&$8$&$120$&$8$&$9$&$2$&$105$&$-4$&$10$&$2$&$128$&$-4$&$15$&$4$&$66$&$8$&$7$&$-1$&$41$&$-1$&$79$&$2$\\
$98$&$8$&$121$&$8$&$11$&$2$&$106$&$-4$&$13$&$2$&$129$&$-4$&$16$&$4$&$68$&$8$&$8$&$-2$&$42$&$1$&$80$&$-2$\\
$101$&$-8$&$124$&$-8$&$12$&$-2$&$107$&$-4$&$15$&$2$&$130$&$-4$&$18$&$4$&$71$&$8$&$9$&$2$&$43$&$1$&$81$&$-2$\\
$104$&$4$&$127$&$4$&$13$&$2$&$108$&$-4$&$17$&$2$&$131$&$-4$&$19$&$4$&$75$&$8$&$11$&$2$&$44$&$2$&$86$&$8$\\
$105$&$-12$&$128$&$-12$&$14$&$2$&$109$&$2\tfrac{2}{3}$&$20$&$2$&$132$&$2\tfrac{2}{3}$&$20$&$4$&$76$&$-8$&$12$&$-1$&$47$&$2$&$88$&$-8$\\
$109$&$8$&$132$&$8$&$16$&$2$&$110$&$-4$&$22$&$-2$&$133$&$-4$&$22$&$-4$&$82$&$4$&$13$&$2$&$48$&$-4$&$94$&$4$\\
$143$&$2$&$159$&$2$&$20$&$2$&$111$&$-1\tfrac{1}{3}$&$24$&$-2$&$134$&$-1\tfrac{1}{3}$&$23$&$-4$&$83$&$-16$&$14$&$2$&$50$&$-4$& & \\
$144$&$4$&$160$&$4$&$21$&$-2$&$112$&$4$&$26$&$2$&$135$&$4$&$25$&$4$&$84$&$16$&$17$&$2$&$52$&$-2$& & \\
$145$&$2$&$161$&$2$&$23$&$-2$&$113$&$4$&$35$&$-2$&$136$&$4$&$29$&$-2$&$85$&$-4$&$19$&$2$&$53$&$2$& & \\
$146$&$4$&$162$&$4$&$27$&$2$&$114$&$-8$&$36$&$2$&$137$&$-8$&$33$&$4$&$86$&$8$&$20$&$2$&$54$&$-4$& & \\
$147$&$-8$&$163$&$-8$&$28$&$2$&$115$&$12$&$37$&$2$&$138$&$12$&$34$&$-4$&$87$&$-8$&$21$&$-2$&$55$&$-4$& & \\
$148$&$-8$&$164$&$-8$&$31$&$-2$&$116$&$-8$&$41$&$-2$&$139$&$-8$&$38$&$4$&$88$&$8$&$24$&$-2$&$56$&$-4$& & \\
$149$&$-8$&$165$&$-8$&$32$&$2$&$117$&$-4$&$42$&$2$&$140$&$-4$&$40$&$-2$&$89$&$8$&$25$&$-2$&$57$&$4$& & \\
$150$&$-4$&$166$&$-4$&$33$&$2$&$118$&$-2\tfrac{2}{3}$&$43$&$2$&$141$&$-2\tfrac{2}{3}$&$45$&$8$&$90$&$16$&$26$&$2$&$58$&$4$& & \\
$151$&$8$&$167$&$8$&$44$&$4$&$119$&$8$&$47$&$4$&$142$&$8$&$46$&$8$&$91$&$-16$&$27$&$2$&$61$&$-4$& & \\
$155$&$12$&$171$&$12$&$48$&$-8$& & &$52$&$-4$& & &$49$&$-8$&$92$&$-8$&$28$&$1$&$62$&$4$& & \\
$156$&$-8$&$172$&$-8$&$62$&$8$& & &$61$&$-8$& & &$51$&$-8$&$93$&$-8$&$29$&$2$&$64$&$4$& & \\
 & & & &$77$&$-4$& & &$67$&$8$& & &$54$&$-8$&$94$&$4$&$30$&$1$&$67$&$4$& & \\
 & & & &$97$&$2$& & &$120$&$2$& & &$55$&$-8$&$95$&$-4$&$31$&$-2$&$69$&$-4$& & \\
 & & & &$98$&$2$& & &$121$&$2$& & &$56$&$8$&$96$&$4$&$32$&$1$&$70$&$4$& & 
 \end{tabular}
\end{ruledtabular}
\end{table*}

%------------------------------------------------------------------
\begin{acknowledgments}
Authors thank Cesnet for providing data management services. KB and KL acknowledge financial support by the Czech Science Foundation under the project No. 16-10042Y. VT acknowledges the Palacky University internal grant No. IGA-PrF-2018-009. KB also acknowledges the financial support of the Polish National Science Center under grant No. DEC-2015/19/B/ST2/01999. All authors acknowledge the projects Nos. LO1305 and CZ.02.1.01./0.0/0.0/16\textunderscore 019/0000754 of the Ministry of Education, Youth and Sports of the Czech Republic.
\end{acknowledgments}

\end{document}